\documentclass[12pt]{iopart}

\usepackage{epsfig}
\usepackage{graphicx}
\usepackage{dcolumn}
\usepackage{bm}

\begin{document}

\title{Investigation of thermoelectric properties of half-metallic Co$_{2}$MnGe by using first principles calculations}
\author{Sonu Sharma and Sudhir K. Pandey}
\address{School of Engineering, Indian Institute of Technology Mandi, Kamand - 175005, India}
\address{Electronic mail: sonusharma@iitmandi.ac.in}

\date{\today}

\begin{abstract}
By combining the electronic structures obtained from first
principles calculations with Boltzmann transport theory we have
investigated the electronic, magnetic and transport properties of
Co$_{2}$MnGe Heusler compound. The density of states plots,
dispersion curves and total energy of paramagnetic and ferromagnetic
(FM) phases clearly show the half-metallic FM ground state for the
compound with an indirect band gap of about 400 meV in the minority
spin channel. It has integer value of the magnetic moment equal to 5
$\mu_{B}$. In the FM phase a very large value ($\sim$550 $\mu$V/K)
of Seebeck coefficient (S) is obtained for down-spin electrons due
to the existence of almost flat conduction band along X to $\Gamma$
direction. The two current model has been used to find the total S
and the obtained value is about 10 $\mu$V/K. The calculated values
of Seebeck coefficient, resistivity and electronic thermal
conductivity show nice agreement with the experimental results.
\end{abstract}

\pacs{75.50.Cc, 74.25.F-, 74.25.fc, 71.20.-b}

\maketitle

\section{Introduction}
Thermoelectric materials are promising for electronic refrigeration
and power generation applications, where heat energy is converted
directly into electricity. Also these materials play an important
role in a global sustainable energy solution. A good thermoelectric
material has high value of the dimensionless figure of merit (ZT),
which is given by\cite{pei,lalonde}
\begin{equation}
{ZT} = S^{2}\sigma T/(\kappa _{e}+\kappa _{l})
\end{equation}
where $S$, $\sigma$, $T$, $\kappa_{e}$ and $\kappa _{l}$  are the
Seebeck coefficient, electrical conductivity, temperature,
electronic thermal conductivity and lattice thermal conductivity of
the material, respectively. Many materials such as narrow band gap
semiconductors, skutterudite, clathrates, complex chalcogenides,
silicides and half-metallic ferromagnets have been identified as
suitable materials for thermoelectric applications\cite{venkat}.
Recently, half-metallic compounds have attracted much interest
because of their exceptional band structures at the Fermi level
($E_{F}$). From the electronic band structure point of view,
half-metals have metallic character for one spin channel and
semiconducting for the other. There exists a high spin-polarization
at the $E_{F}$ and many Heusler alloys belong to this family. The
very first time half-metallic ferromagnetism has been introduced by
Groot et al.\cite{groot} in the NiMnSb half-Heusler alloy. After
that many Heusler alloys were found to be half-metallic in different
theoretical\cite{igalanakis,kubler,ishida,fujii,picozzi,barth,candan}
and experimental\cite{hanseen,kirillova} studies.

For the spin-polarized materials the total Seebeck coefficient can
be evaluated by using two current model\cite{xiang,botana}.
According to this model the expression for total Seebeck coefficient
is given by formula:
\begin{equation}
S =
[\sigma(\uparrow)S(\uparrow)+\sigma(\downarrow)S(\downarrow)]/[\sigma(\uparrow)+\sigma(\downarrow)]
\end{equation}
where, S and $\sigma$ are Seebeck coefficient and electrical
conductivity for both the spin channels. Here, it is important to
note that the value of S for semiconductors is found to be about 100
times larger than that of the metals\cite{wood}. As the
half-metallic compounds show metallic behaviour for one spin channel
and semiconducting for the other. The total thermopower is expected
to be more in these compounds than that of the metals and less than
the semiconductors. Experimental studies also suggest that the
half-metallic ferromagnetic materials have relatively larger value
of the Seebeck coefficient\cite{bbalke,aliev}. Co$_{2}$MnX-like
Heusler alloys are expected to be good thermoelectric materials in
the high temperature range because there exits a band gap in the
minority spin state of these alloys\cite{kandpal} and they have high
Curie temperature\cite{brown}. Also it is possible to tune their
magnetic properties\cite{kubler} and $E_{F}$ position\cite{balke} by
the substitution of atoms that have similar radii and charges.

For many years, it was very challenging task to calculate accurately
the transport properties of materials\cite{nag}. Recently, progress
has been made by combining the first principles band structure
calculations and the Boltzmann transport
theory\cite{djsingh,madsen,singh}. According to this theory
electrical conductivity can be calculated from the electronic
structures\cite{singh,mizutani} and usually the electrical
conductivity has large dependence on the density of
states\cite{yang}. Co$_{2}$MnGe has a narrow band gap in the
minority spin state, so it is expected that minority spin state will
give large value of S than that of the metals. The Curie temperature of the compound is
nearly 905 K\cite{bbalke}. In the present work we have studied the
electrical transport properties of Co$_{2}$MnGe in both the
paramagnetic and FM phases for high temperature range.

The full-potential linearized augmented-plane wave methods within
local spin density approximation (LSDA) has been employed to
calculate the electronic band structures, magnetic states and
electronic transport properties of Co$_{2}$MnGe. The study of
electronic and magnetic properties shows that this compound is
half-metallic ferromagnet with the band gap of about 400 meV in the
minority spin channel. The temperature dependent electronic
transport coefficients are evaluated for both the paramagnetic (PM)
and FM phases of the compound. The negative value of S for both
phases indicates the presence of N-type charge carriers. In the FM
phase the obtained value of S for down spin electrons is $\sim$550
$\mu$V/K, which is almost 55 times greater than that of the up spin
electrons. The  total S is computed by using two current model and
its value is found to be about 10 $\mu$V/K. The calculated values of
S, $\rho$ and $\kappa_{e}$ are dominated by majority spin electrons
and show fairly good agreement with the available experimental data.

\section{Computational details and Crystal structure}
The calculations of electronic and magnetic properties have been
performed by using the full-potential linearized augmented
plane-wave (FP-LAPW) method within the density functional theory
(DFT) implemented in WIEN2k code\cite{blaha}. The
exchange-correlation functional has been treated within the local
spin density approximation (LSDA) of Perdew and Wang \cite{perdew}.
The muffin-tin radii were set to 2.2 Bohr for Co, Mn and Ge atoms.
$R_{MT}K_{MAX}$ which determines the matrix size for convergence was
set equal to 7, where $R_{MT}$ is the smallest atomic sphere radii
and $K_{max}$ is the plane wave cut-off. The maximum \textit{l}
value ($l_{max}$) for partial waves used inside atomic spheres was
kept 10. The experimental lattice parameter of 5.743
\AA\cite{webster} has been used in the calculations. The volume of the unit cell was about 284 Bohr$^{3}$. The self
consistent loop was repeated until the total energy/cell of the
system converges to less than 10$^{-4}$ Ry. The $40\times 40\times
40$ k-point mesh is used as the accurate calculations of transport
properties of metals require the dense sampling of the Brillouin
zone. The transport properties of the compound have been calculated
by using the BolzTrap code\cite{singh} under the constant relaxation
time approximation for the charge carriers. The Fermi energy obtained in the self consistent calculations was considered as the chemical potential for the calculation of transport properties of the compound. The values of chemical potential used in the calculation were 0.7926 eV and 0.8051 eV for PM and FM phases, respectively.

Full Heusler compound Co$_{2}$MnGe crystallizes in $L2_{1}$ crystal
structure with space group $Fm$-$3m$. Co atoms are placed on the
Wyckoff position 8c (1/4, 1/4, 1/4). Mn and Ge atoms are located at
the Wyckoff position 4a (0, 0, 0) and 4b (1/2, 1/2, 1/2),
respectively\cite{brown}. The cubic $L2_{1}$ structure consists of
four interpenetrating fcc sub-lattices, two of which are equally
occupied by Co atoms. The two Co-site fcc sub-lattices, combine to
form a simple cubic sub-lattice. The Mn and Ge atoms occupy
alternatively, the centre of the simple cubic sub-lattice resulting
in a CsCl-type superstructure\cite{kandpal}.

\section{Results and discussions}
\subsection{Paramagnetic Phase}

The total density of states (TDOS) and partial density of states
(PDOS) plots for the PM phase are shown in Fig. 1. The $E_{F}$ is
represented by dashed line. In Fig. 1(a) the total DOS at $E_{F}$ is
$\sim$7 states/eV/f.u. (f.u. $\equiv$ formula unit) for both the
spin channels, which is very large. Such a large value of TDOS at
the $E_{F}$ may be consider as indication of the FM ground state as
per the Stoner theory\cite{skpandey}. It is also clear from the
figure that the antibonding bands are extended up to about 0.6 eV
below $E_{F}$. Based on Stoner theory one can easily guess that the
total energy of the system will be minimized if there is shifting in
spin-up and spin-down bands by about 0.6 eV below and above the
$E_{F}$, respectively. It give rise to half-metallic FM ground state
for the compound. This is also evident from the total energy
calculations, where the total energy of FM phase is about 1.1 eV
less than the PM phase. From the PDOS of Co atom in Fig. 1(b), it is
clear that the contribution of \textit{t$_{2g}$} and
\textit{e$_{g}$} states at the $E_{F}$ is $\sim$0.4 states/eV/atom
and $\sim$1.3 states/eV/atom, respectively, in both the spin
channels. It is also evident from Fig. 1(c) that the Mn
\textit{t$_{2g}$} and \textit{e$_{g}$} PDOS at $E_{F}$ is $\sim$1
and 2 states/eV/atom, respectively. The PDOS of Ge atom (Fig. 1(d))
shows that the occupancy at the $E_{F}$ for p-orbital is $\sim$0.5
states/eV/atom with very small contribution of \textit{s} and
\textit{d}-orbitals. From these figures it is clear that both
\textit{t$_{2g}$} and \textit{e$_{g}$} orbitals are mainly
contributing to the total DOS at the $E_{F}$. These orbitals are
expected to be responsible for the FM ground state.

The dispersion curves along the high symmetry directions of the
first Brillouin zone are presented in Fig. 2(a). It is evident from
the figure that bands labeled by 1, 2 and 3 are lying just above the
$E_{F}$ whereas bands 6, 7 and 8 are lying just below it. Bands 4
and 5 cross the $E_{F}$ at 8 different k-points. The bands 2-7 are
concentrated around W-point in the energy range of about -0.2 to 0.2
eV. Shifting of these bands can minimize the total energy of the
system and may lead to the FM ground state for this compound.

In Fig. 2(b-e) temperature induced carrier concentration per unit cell, calculated Seebeck coefficient (S), electrical
conductivity ($\sigma/\tau$) and electronic thermal conductivity
($\kappa_e$/$\tau$), where $\tau$ is the relaxation time, are
displayed for the temperature ranging from 900 K to 1200 K. It is evident from Fig. 2(b) that the temperature induced charge carriers show linear variation with temperature and the absolute value of N decreases from $\sim$ 0.36 e/unit cell at 900 K to $\sim$ 0.007 e/unit cell at 1200 K.
Fig. 2(c) shows the linear temperature dependence of the S. The value of
S is negative for the entire temperature range, which indicates
N-type of charge carriers in the PM phase of the compound. The value
of S increases from $\sim$16 $\mu$V/K at 900 K to $\sim$26 $\mu$V/K
at 1200 K. Fig. 2(d and e) represents the temperature variation
of the $\sigma/\tau$ and $\kappa_e$/$\tau$. $\sigma/\tau$ shows the
slight deviation from the linear temperature dependence, whereas
$\kappa_e$/$\tau$ varies almost linearly with temperature. The value
of $\sigma/\tau$ decreases from about 2.43 $\times
10^{20}(\Omega^{-1}m^{-1}s^{-1}$) at 900 K to about 2.41 $\times
10^{20}(\Omega^{-1}m^{-1}s^{-1}$) at 1200 K. The value of
$\kappa_e$/$\tau$ increases from $\sim$0.52 $\times 10^{16} (W
m^{-1}K^{-1}s^{-1}$) at 900 K to $\sim$0.72 $\times 10^{16} (W
m^{-1}K^{-1}s^{-1}$) at 1200 K.

\subsection{Ferromagnetic Phase}
The TDOS and PDOS for the FM solution are shown in Fig. 3. From the
TDOS plots (Fig. 3(a)), it is clear that the majority spin states
are occupied at the $E_{F}$ with occupancy of 1.09 states/eV/f.u.
and minority spin states are empty. Thus for spin-up channel this
compound is metallic and for spin-down channel it shows
semiconducting behaviour. On comparing Fig. 1(a) and 3(a), one can
find that in spin-up channel \textit{t$_{2g}$} and \textit{e$_{g}$}
states shifts towards lower energy whereas, in spin-down channel
these states shift towards higher energy as conjectured above.
Because of this shifting the almost empty region of spin-down
channel about 0.6 eV below the $E_{F}$ for PM phase, is exactly
lying on the $E_{F}$ in FM phase. Therefore there is creation of
band gap in the minority spin channel. This rigid band shift appears
to be responsible for the FM ground state in this compound. The Fig.
3 (b-d) show the PDOS plots for the Co, Mn and Ge atoms. In Fig. 3
(b) the PDOS plots for Co 4\textit{s}, \textit{t$_{2g}$} and
\textit{e$_{g}$} orbitals are shown. It is clear from the figure
that in case of spin-up channel, the Co \textit{t$_{2g}$} and
\textit{e$_{g}$} states are occupied and for spin-down channel these
states are unoccupied. The occupancy at $E_{F}$ for
\textit{t$_{2g}$} and \textit{e$_{g}$} states are 0.2 states/eV/atom
and 0.07 states/eV/atom, respectively. One can also see the
negligible contribution of the other states. The PDOS plot of
4\textit{s}, \textit{t$_{2g}$} and \textit{e$_{g}$} bands of Mn atom
is presented in Fig. 3(c). The \textit{t$_{2g}$} states mainly
contribute ($\sim$0.4 states/eV/atom) at the $E_{F}$ with negligibly
small contribution from other states. This result is quite different
from the PM phase where both \textit{t$_{2g}$} and \textit{e$_{g}$}
orbitals contribute to the $E_{F}$. In Fig. 3(d) the
density of states for 4\textit{s}, 4\textit{p} and
3\textit{d}-orbitals of Ge atoms are presented which shows
negligibly small contribution of these states at the $E_{F}$ in
comparison to \textit{t$_{2g}$} state of Co and Mn atoms. This
result clearly suggests that \textit{t$_{2g}$} electrons of Mn and
Co atoms are responsible for the transport behaviour of the
compound.

The dispersion curves are presented in Fig. 4 along the high
symmetry directions of the first Brillouin zone. These dispersion
curves also show the half-metallic ground state of this compound.
For spin-up channel in Fig. 4(a), it is clear that bands 1, 2 and 3
cross the $E_{F}$ at 5 k-points and are responsible for the transport
properties of this compound. The bands which are not crossing the $E_{F}$ will contribute negligibly
small to the transport properties. The electrons lying in these bands have energy
more than $\sim$1 eV which corresponds to $\sim$12000 K temperature and are not going to influence the transport properties of the compound in the temperature range studied here. Also the shift in chemical potential is about 1 meV which is negligible in comparison to 1 eV mentioned above. Therefore bands crossing the $E_{F}$ will mainly contribute in the transport properties of the compound.
 Bands 4 to 8 have shifted deep into the
valence band (VB). For spin-down channel bands 1 to 7 have shifted
far away from $E_{F}$ into the conduction band (CB) and there is a
clear cut gap between bands 7 and 8, as indicated in Fig. 4(b). This
shifting appears to be mainly responsible for the FM ground state
for this compound as stated earlier. The VB maximum is at
$\Gamma$-point and CB minimum is at X-point, therefore there exists
an indirect band gap in the spin-down channel. The computed energy
gap with LSDA exchange correlation functional is about 0.40 eV which
is smaller than 0.54 eV\cite{picozzi} energy gap obtained by using
GGA exchange correlation functional. However, Galanakis et
al.\cite{galanakis} have observed no real gap in this full-Heusler
alloy, by using the full-potential screened Korringa-Kohn-Rostoker
(FSKKR) Green’s function method in conjunction with the local spin
density approximation. There are three degenerate bands at
$\Gamma$-point. Along $\Gamma$ to L and $\Gamma$ to X direction the
degeneracy is partially lifted. The effective mass of the VB is
expected to have much smaller value in comparison to the CB as there
exits almost flat band in CB along X to $\Gamma$ direction. The
large effective mass for CB is expected to give a very high and
negative value of thermopower from the minority spin channel.

The total magnetic moment per formula unit for this compound is 5.0
$\mu_{B}$ with the contribution from Co, Mn, Ge and interstitial
region is 1.06, 2.84, -0.02 and 0.05 $\mu_{B}$, respectively. The
total magnetic moment is in agreement with the experimental value
\cite{ambrose}. The Mn atom carries the largest magnetic moment and
ferromagnetically coupled with the Co atom. Although the magnetic
moment of Ge is small but it is coupled antiferromagnetically with
Co and Mn atoms. The integer value of total magnetic moment confirms
its half metallic character.

The transport coefficients are normally found to be very much
sensitive to the k-point sampling of the Brillouin zone. In order to
check the k-points dependence of the transport coefficients we have
calculated the Seebeck coefficient (S), electrical conductivity
($\sigma/\tau$) and electronic thermal conductivity
($\kappa_e$/$\tau$) at different k-points for spin-up channel. These
coefficients correspond to 64000, 125000 and 175616 k-points in the
full Brillouin zone are shown in Fig. 5. It is evident from the
figure that at low temperature S and $\sigma/\tau$ are very much
sensitive to the k-points considered in the calculations whereas
$\kappa_e$/$\tau$ does not show any significant k-point dependence.
The values of S and $\sigma/\tau$ above 150 K and 200 K are found to
be almost the same for all k-points. Thus the transport coefficients
data below 150 K may not be very much reliable. In order to evaluate
the low temperature transport behaviour of the compound a careful
study is required.

The temperature induced carrier concentration per unit cell, Seebeck coefficient, electrical conductivity and electronic
thermal conductivity versus temperature plots corresponding to both
the spin channels are presented in Fig. 6. It is clear from Fig. 6(a) that the value of N increases from $\sim$ 0.001 e/unit cell at 150 K to $\sim$ 0.01 e/unit cell at 900 K for spin-up channel. Fig. 6(b) shows that for spin-down channel the value of N increases from about 10$^{-6}$ e/unit cell at 150 K to about 0.03 e/unit cell at 900 K. From Fig. 6(c and d), it
is clear that the obtained value of Seebeck coefficient is negative
for the entire temperature range and suggesting the presence of
N-type charge carriers. The absolute value of S increases from
$\sim$4$\mu$V/K at 150 k to $\sim$26$\mu$V/K at 900 K for spin-up
channel and decreases from $\sim$900$\mu$V/K at 150 K to
$\sim$274$\mu$V/K at 900 K for down spin electrons. At room
temperature the obtained values of S are nearly 10 $\mu$V/K and 550
$\mu$V/K, for spin-up and spin-down channels, respectively. Such a
large value of S for spin-down electrons is essentially due to the
presence of the almost flat CB along X to $\Gamma$ direction, as
seen in Fig. 4(b). The electrical conductivities for spin-up channel
is shown in Fig. 6(e). It is clear from the figure that the
conductivity firstly increases from $\sim$4.73$\times
10^{20}(\Omega^{-1}m^{-1}s^{-1}$) at 150 K to $\sim$4.75$\times
10^{20}(\Omega^{-1}m^{-1}s^{-1}$) at 300 K and then decreases slowly
with temperature to the value $\sim$4.70$\times
10^{20}(\Omega^{-1}m^{-1}s^{-1}$) at 900 K. At room temperature the
value of $\sigma/\tau$ is about 4.75$\times
10^{20}(\Omega^{-1}m^{-1}s^{-1}$). The electrical conductivity of
spin-down channel is shown in Fig. 6(f). The conductivity increases
with increase in temperature from $\sim$2$\times
10^{15}(\Omega^{-1}m^{-1}s^{-1}$) at 150 K to $\sim$0.05$\times
10^{20}(\Omega^{-1}m^{-1}s^{-1}$) at 900 K and this compound have
semiconductor like behaviour for down-spin electrons. At room
temperature the value of $\sigma/\tau$ is nearly 0.001 $\times
10^{20}(\Omega^{-1}m^{-1}s^{-1}$) and is very low in comparison to
the up spin electrons. The electronic thermal conductivities for
both the spin channels are shown in Fig. 6(g and h). For spin-up
electrons the thermal conductivity increases linearly from
$\sim$0.17$\times 10^{16}(W m^{-1}K^{-1}s^{-1}$) at 150 K to
$\sim$1$\times 10^{16}(W m^{-1}K^{-1}s^{-1}$) at 900 K. In case
of down-spin electrons the thermal conductivity shows non linear
variation and increases from $\sim$2$\times 10^{11}(W
m^{-1}K^{-1}s^{-1}$) at 150 K to $\sim$0.04$\times 10^{16}(W
m^{-1}K^{-1}s^{-1}$) at 900 K. The room temperature values of
$\kappa_e$/$\tau$ are about 0.34 $\times 10^{16}(W
m^{-1}K^{-1}s^{-1}$) and 0.001 $\times 10^{16}(W
m^{-1}K^{-1}s^{-1}$) for spin-up and spin-down electrons,
respectively. For both the spins $\kappa_e$/$\tau$ increases with
increase in temperature.

In order to calculate the total S one can rewrite the equation (2)
as,
 \begin{equation}
S=S(\uparrow)\{
\frac{1+[S(\downarrow)/S(\uparrow)][\sigma(\downarrow)/(\sigma(\uparrow)]}{1+[\sigma(\downarrow)/\sigma(\uparrow)]}\}
\end{equation}
 Using the room temperature values of S and $\sigma$ in this equation, we find that the ratio of S($\downarrow$) to S($\uparrow$) is about 55, whereas the ratio of $\sigma(\downarrow)$ to $\sigma(\uparrow)$ is about 0.0001. Due to the extremely small value of $\sigma(\downarrow)$/($\sigma(\uparrow)$] in comparison to  S($\downarrow)$/S$(\uparrow)$ the total S is essentially dominated by the value of S$(\uparrow)$ only. The total S obtained by using  equation (3) is presented in Fig. 7. It is evident from the figure that the total S is negative and varies linearly with temperature. The value of S increases from $\sim$4$\mu$V/K at 150 K to $\sim$29$\mu$V/K at 900 K. The calculated value of S at room temperature is nearly 10$\mu$V$K^{-1}$ and is closer to the experimental values\cite{bbalke,ouardi}. At low temperature $\sigma/\tau$ and $\kappa_e$/$\tau$ are mainly contributed by the spin-up channel as evident from Fig. 6(c-f). We have evaluated the $\rho$ and $\kappa_{e}$ at room temperature for spin-up channel by using the value of $\tau=0.5\times10^{-14}$, because the room temperature value of $\tau$ is typically 10$^{-14}$ to $10^{-15}$ seconds\cite{ashcroft}. The calculated values of $\rho$ and $\kappa_{e}$ are 0.42 $\mu \Omega m$  and 17.5 $Wm^{-1}K^{-1}$, respectively, which are in good agreement with the experimental values\cite{ouardi}.

\section{Conclusions}
The electronic, magnetic and transport properties of Co$_{2}$MnGe
have been investigated by combining the electronic structures
calculated from first principles methods with the Boltzmann
transport theory. The paramagnetic and ferromagnetic density of
states plots, electronic band structures and total energy
calculations clearly suggest the half-metallic ferromagnetic ground
state for the compound. The indirect band gap from $\Gamma$ to X
points is found to be $\sim$400 meV in the spin-down channel. The
total magnetic moment obtained from the calculations is 5 $\mu _{B}$
per formula unit. The integral value of magnetic moment is in
accordance with the half-metallic ground state. For spin-down
channel the value of S is about 55 times greater than that of the
spin-up channel and conductivity is about 0.0001 times smaller. The
large value of Seebeck coefficient in the spin-down channel is
attributed to the presence of almost flat CB along $\Gamma$ to X
direction. By using two current model the value of total Seebeck
coefficient comes out to be $\sim$10 $\mu$V/K and only spin-up
channel is found to be responsible for the transport behaviour of
the compound. The values of Seebeck coefficient, resistivity and
electronic thermal conductivity show fairly good agreement the
experimental values.



\section{Figures}

\begin{figure}
  \includegraphics[width=14cm]{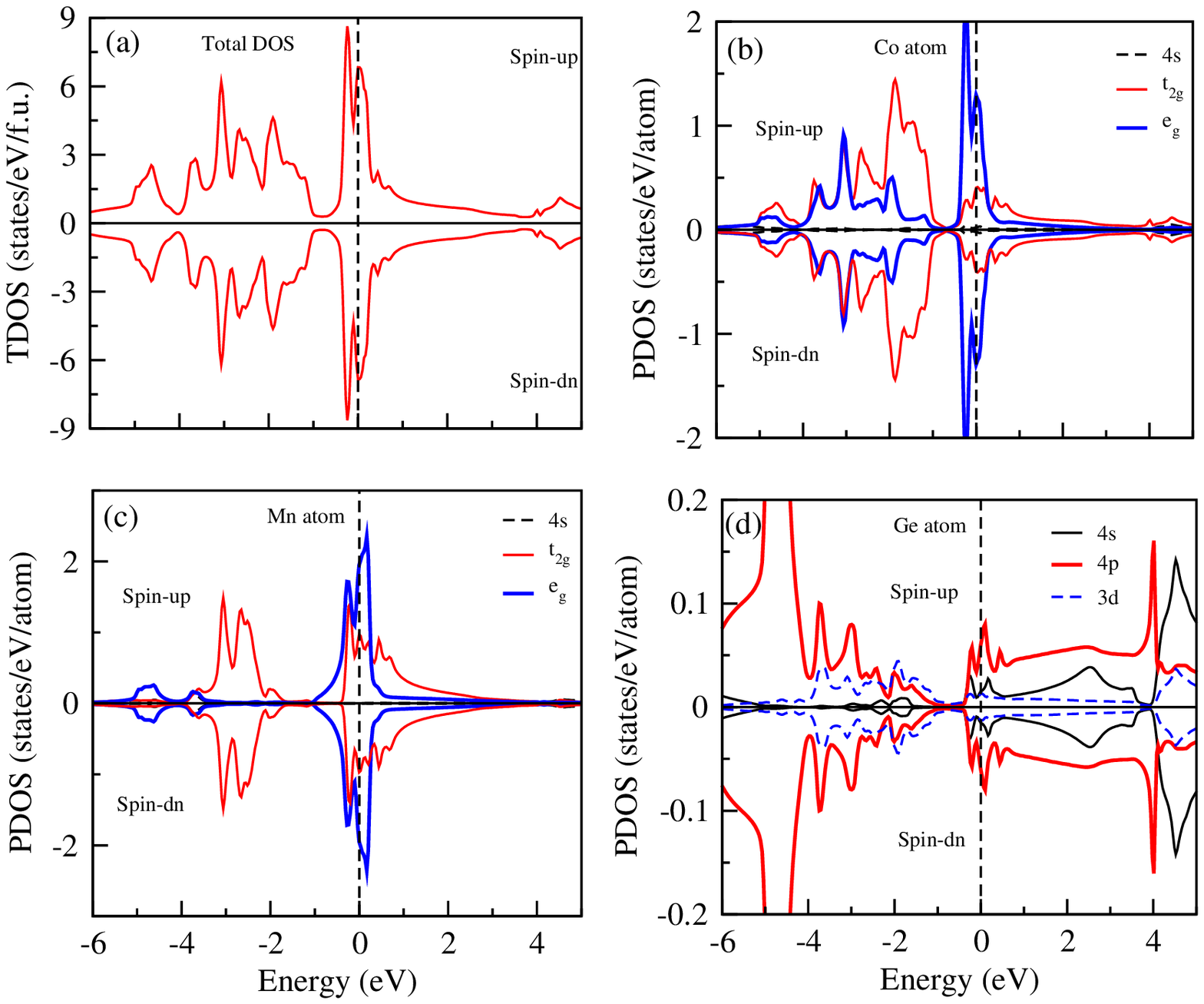}\\
  \caption{(Color online) Total and partial density of states plots for
Co$_{2}$MnGe in the paramagnetic phase. Shown are (a) the TDOS plot,
(b) PDOS of Co atom, (c) PDOS of Mn atom and (d) PDOS of Ge
atom.}\label{Fig1}
\end{figure}

\begin{figure}
  \includegraphics[width=14cm]{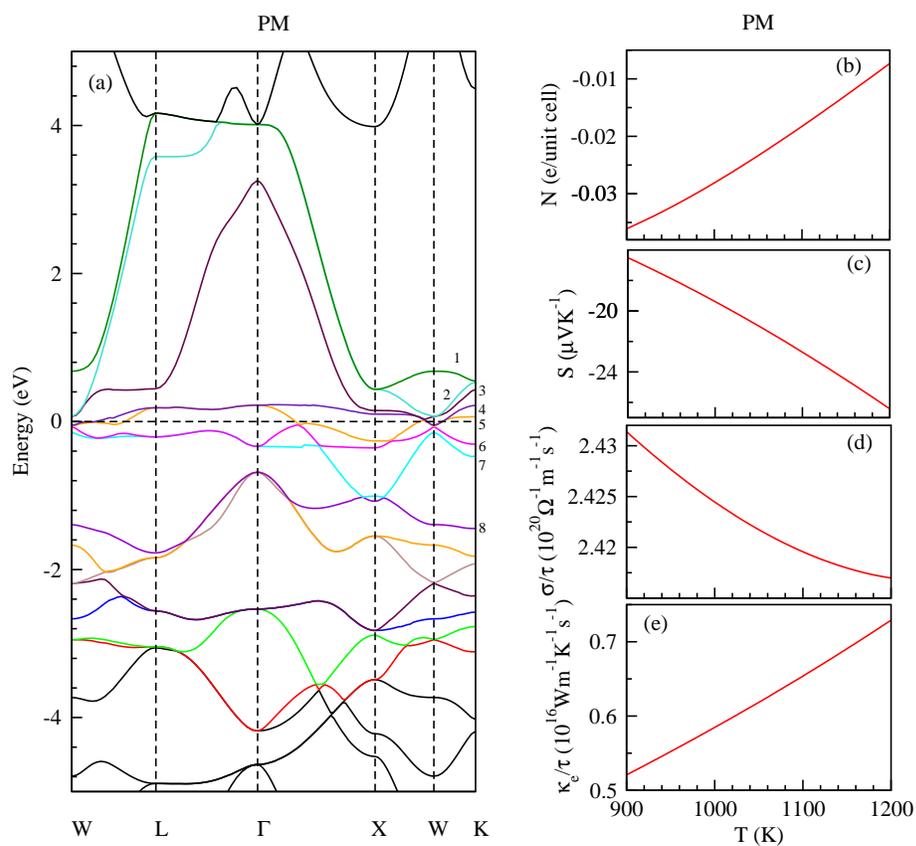}\\
  \caption{(Color online) Electronic band structure and transport
properties of paramagnetic Co$_{2}$MnGe. Shown are the (a)
electronic band structure, (b) temperature induced carrier
concentration per unit cell. Temperature variation of Seebeck
coefficient, Electrical conductivity and Electronic thermal
conductivity are shown in (c), (d) and (e),
respectively.}\label{Fig2}
\end{figure}

\begin{figure}
  \includegraphics[width=14cm]{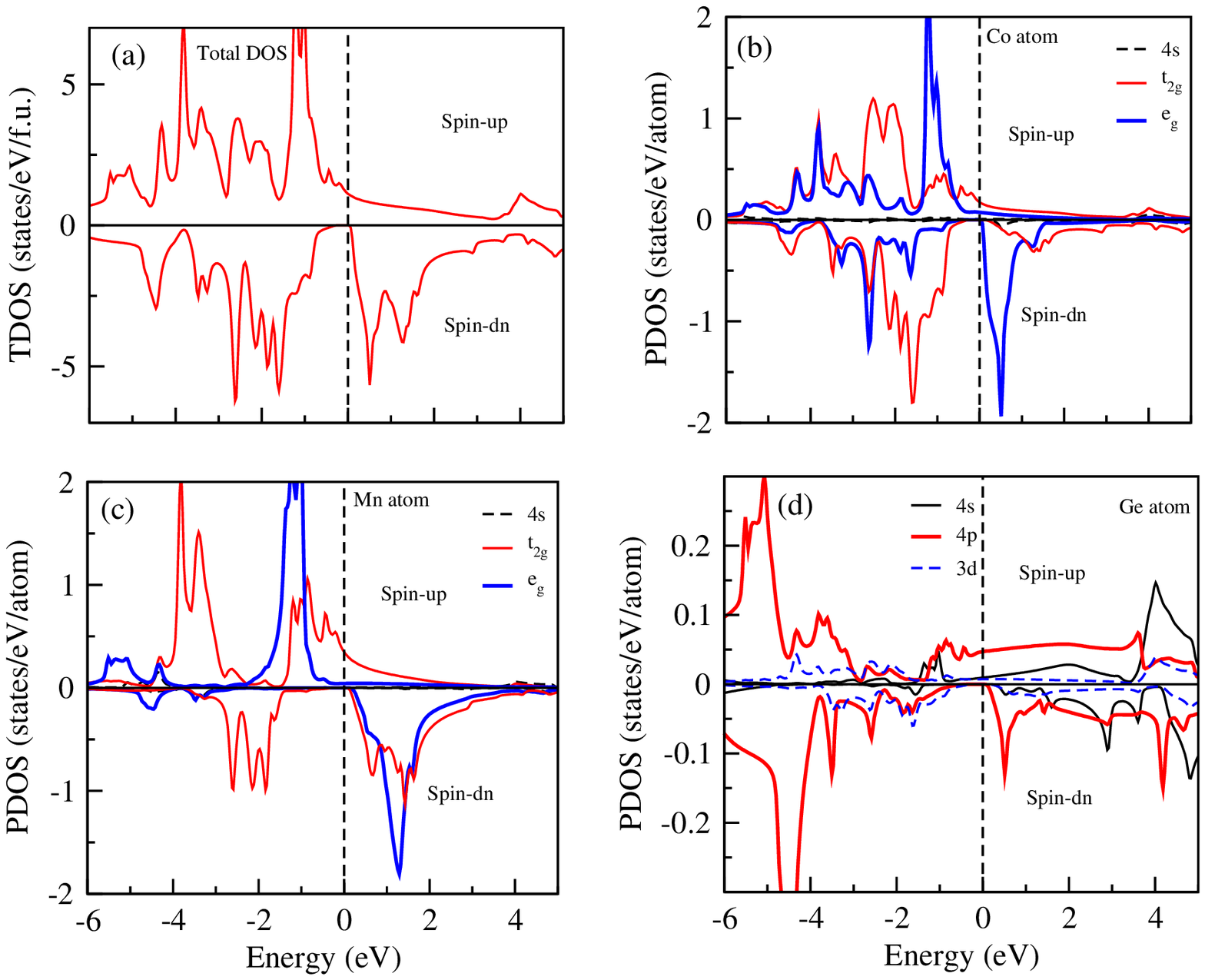}\\
  \caption{(Color online) Total and partial density of states plots for
Co$_{2}$MnGe in the ferromagnetic phase. Shown are (a) the TDOS
plot, (b) PDOS of Co atom, (c) PDOS of Mn atom and (d) PDOS of Ge
atom.}\label{Fig3}
\end{figure}

\begin{figure}
  \includegraphics[width=14cm]{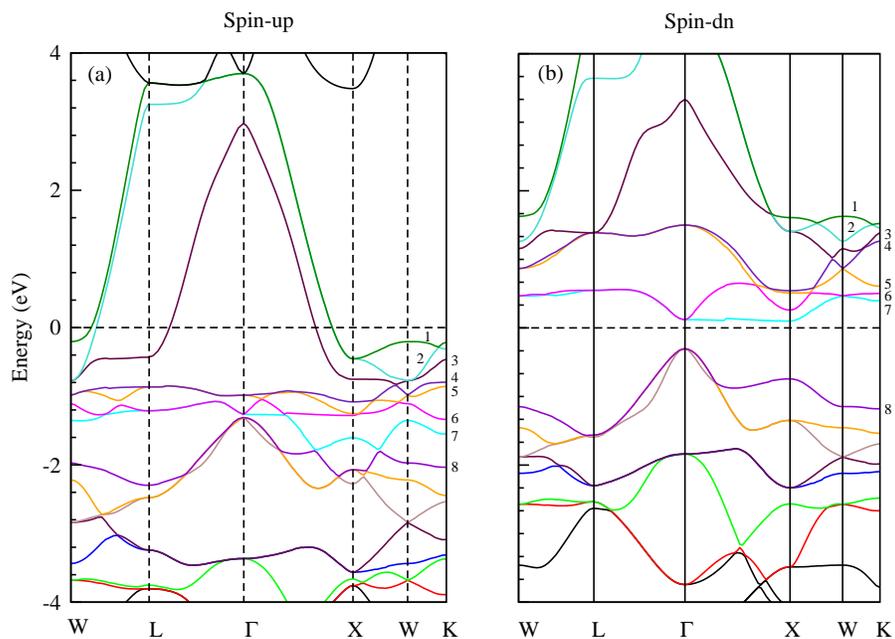}\\
  \caption{(Color online) Electronic band structures of ferromagnetic
Co$_{2}$MnGe (a) spin-up channel and (b) spin-down
channel.}\label{Fig4}
\end{figure}

\begin{figure}
  \includegraphics[width=14cm]{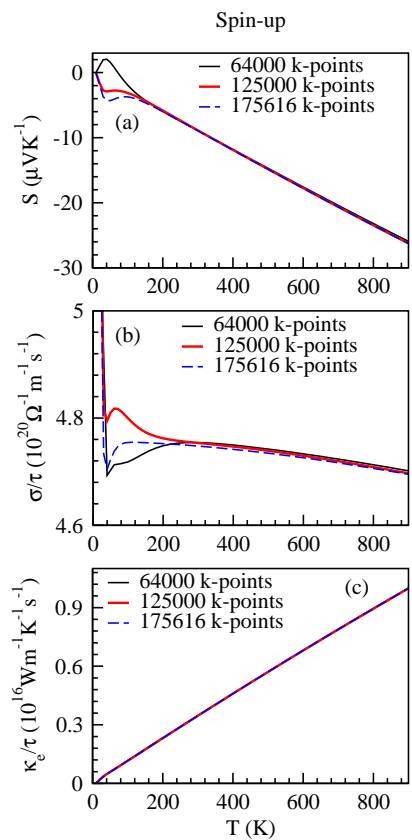}\\
  \caption{(Color online) k-point dependence of (a) Seebeck coefficient, (b) Electrical conductivity and (c) Electronic thermal
conductivity.}\label{Fig5}
\end{figure}

\begin{figure}
  \includegraphics[width=14cm]{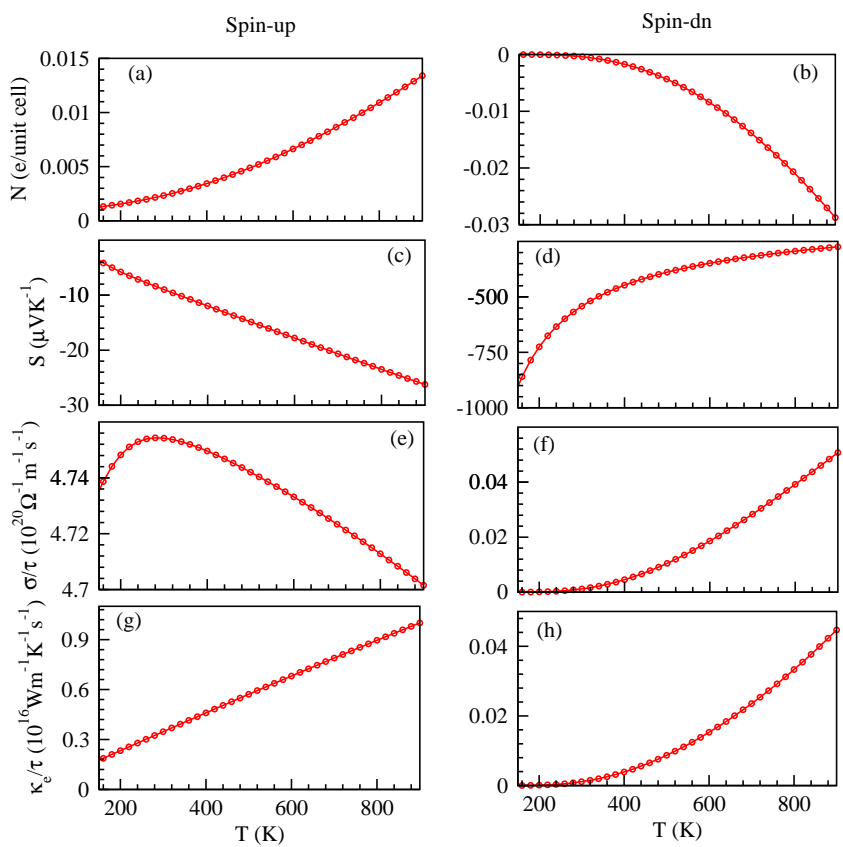}\\
  \caption{(Color online) Variation of the transport properties in
ferromagnetic phase with temperature for spin-up and spin-down
channels. Shown are (a and b) temperature induced carrier concentration per unit cell, (c and d) Seebeck coefficient, (e and f)
Electrical conductivity and (g and h) Electronic thermal
conductivity.}\label{Fig6}
\end{figure}

\begin{figure}
  \includegraphics[width=14cm]{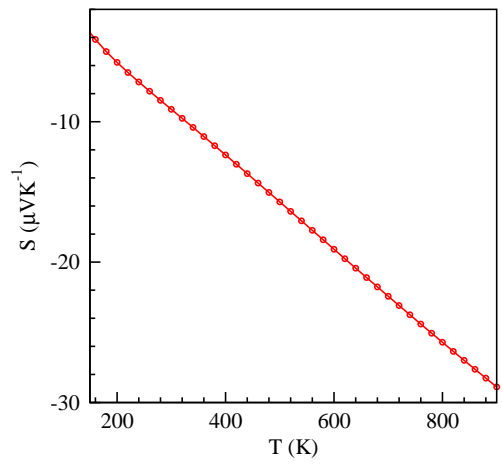}\\
  \caption{(Color online) Variation of total Seebeck coefficient (S)
with temperature.}\label{Fig7}
\end{figure}

\end{document}